\documentclass[conference]{IEEEtran}
\usepackage{cite}

\ifCLASSINFOpdf
\usepackage[pdftex]{graphicx}
\graphicspath{{../pdf/}{../jpeg/}}
\DeclareGraphicsExtensions{.pdf,.jpeg,.png}
\else
\usepackage[dvips]{graphicx}
\graphicspath{{../eps/}}
\DeclareGraphicsExtensions{.eps}
\fi

\usepackage{amsmath}
\usepackage{amsfonts}
\usepackage{amssymb}
\usepackage{amsthm}
\usepackage{array}
\usepackage[caption=false,font=footnotesize]{subfig}
\usepackage{dblfloatfix}
\usepackage{tikz}
\usetikzlibrary{shapes,fit}
\usepackage{pgfplots}
\usepackage{xcolor}
\usepackage[outline]{contour}
\contourlength{3pt}

\usepackage{multirow}
\usepackage{threeparttable}
\usepackage{hyperref}
\usepackage{xfrac}
\usepackage{ulem}
\newtheorem{thm}{Theorem}

\newtheorem{remark}{Remark}
\newtheorem{definition}{Definiton}

\begin{document}
%
\title{Precoding for a Class of Peak-Constrained Dirty Paper Channels with a Discrete State}
\author{
\IEEEauthorblockN{Zhenyu (Charlus) Zhang, Anas Chaaban}
\IEEEauthorblockA{School of Engineering, University of British Columbia, Kelowna, V1V1V7 BC, Canada\\
{zhenyu.zhang@alumni.ubc.ca}, {anas.chaaban@ubc.ca}}
}

\maketitle

\begin{abstract}
	The dirty paper channel (DPC) under a peak amplitude constraint arises in an optical wireless broadcast channel (BC), where the state at one receiver is the transmitted signal intended for the other receiver(s). This paper studies a class of peak-constrained DPC that is applicable to the optical wireless BC, where the channel state (i.e, `dirt') takes values from some evenly-spaced grid. For the discrete-state DPC studied this paper, a capacity upper bound is obtained from its state-free counterpart. To lower bound its capacity, classical dirty paper coding schemes are revisited, including Costa's coding for DPC and Tomlinson-Harashima (TH) precoding, which serves as benchmark schemes. To improve the benchmark performance, two new precoding schemes are proposed for the discrete-state DPC. Although the proposed schemes do not achieve the state-free capacity contrary to what is known about the Costa's DPC, achievable rates within a small gap to the state-free capacity are demonstrated for the discrete-state DPC. Using the proposed precoding scheme in a two-user peak-constrained Gaussian BC, a new capacity inner bound (IB) is obtained, and is shown to outperform the truncated Gaussian (TG) based IB and is comparable to the best-known IB.
\end{abstract}


%
\IEEEpeerreviewmaketitle

\section{Introduction} \label{sec:intro}
In optical wireless communication (OWC), the following interference arises: Intersymbol interference (ISI) in multipath transmission scenarios, intercolor interference (ICI) in multicolor transmission scenarios, and interuser interference (IUI) in broadcasting scenarios. All the interference types share a common feature that the interference is known at the transmitter but not to receiver. From an information thoery perspective, the \emph{channel with state}, can well model an OWC link disturbed by ISI, ICI, or IUI, where the interference is modeled as the channel state \cite[Sec. 7.6]{book-NIT}. 

To understand the importance of channel with state, let us look at the more-developed radio-frequency (RF) communication channel as an example, where the link with known interference at the transmitter is modeled as a power-constrained Gaussian channel with state (also known as a dirty paper channel, DPC \cite{Costa1983}). The milestone work by Costa in 1983 \cite{Costa1983} states that the power-constrained DPC has the same capacity of a Gaussian channel without state. The proof is based on the Gelfand-Pinsker Theorem \cite[Theorem 7.3]{book-NIT} which states the capacity of general dirty paper channels. Costa's result has motivated code design for the RF DPC \cite{erez2005capacity}, and enabled developing capacity-achieving coding schemes for broadcast channels (BCs) based on the DPC for both the scalar \cite{yu2001trellis} and the vector cases \cite{yu2005trellis,weingarten2006capacity,mohseni2006capacity}.

Different from RF communication, OWC DPC is constrained by a peak amplitude constraint instead of a power constraint due to physical properties of optical emitters, such as LEDs or laser diodes \cite{Lapidoth-Moser-Wigger-2009,chaaban2016capacity}. In spite of the importance of the peak-constrained DPC in OWC, little is known about the peak-constrained DPC in the current literature. Note that, for a peak-constrained Gaussian channel, the optimal code is shown to follow a discrete distribution \cite{smith1971information,chan2005capacity}, instead of the Gaussian distribution that is optimal in the power-constrained Gaussian channel. \emph{The code design for the OWC DPC is likely different from that for the RF DPC.} Hence, investigating the capacity of the OWC DPC is an interesting research direction and is open in the literature.

Hence, we are motivated to study the case when the DPC-based broadcasting is adopted together with discrete signaling in OWC. We propose that the alphabets that help to encode messages of different users should be drawn from the same grid whose elements are evenly spaced. Then, a class of DPC with a discrete state will arise in the proposed setting. We examine the concept in the peak-constrained Gaussian BC with a scalar input first in this paper, and leave the investigation for the vector-input case to future work.

In this paper, we study the capacity of the peak-constrained discrete-state dirty paper channel (DPC) that arises in optical wireless broadcasting with discrete signaling. To this end, we first revisit the classical coding scheme from the power-constrained DPC, including Costa's coding for DPC and Tomlinson-Harashima (TH) precoding as benchmarks. To improve the performance obtained by these classical schemes, we then propose two new precoding schemes for the peak-constrained discrete-state DPC, and all are based on the modulo operation. It turned out that the achievable rate of Costa's coding, TH precoding, and the first proposed scheme can only be evaluated through their lower bounds, however, the second proposed scheme can be approximated through the approximation of the entropy over the Gaussian-mixture distributions. Numerical results show that both the proposed precoding schemes outperforms the benchmark schemes. We further examine the application of the peak-constrained discrete-state DPC in a two-user peak-constrained Gaussian BC with a scalar input. 
By applying Scheme 2 to the two-user peak-constrained Gaussian broadcast channel (BC), a new capacity inner bound (IB) is obtained, which is shown to outperform the capacity IB that is based on the truncated Gaussian (TG) distribution and performs comparable to the best known inner bound that is based on a discrete distribution. Thus, the application of the peak-constrained discrete-state DPC in a vector BC becomes promising. 

In the rest of this paper, after defining some notations, Sec. \ref{sec:pre} provides basics for BC and define the discrete-state DPC that arises to facilitate the discussion and analysis in what follows. Then, Sec. \ref{sec:r0} revisit the classical coding schemes of the power-constrained DPC in the peak-constrained counterpart to form benchmark performance. Then, Sec. \ref{sec:newR} introduces two new precoding schemes. The achievable rates of all precoding schemes are numerically compared in Sec. \ref{sec:figures}, and a new capacity inner bound for the peak-constrained BC has also been shown in this section. Finally, Sec. \ref{sec:concl} concludes this paper.

\emph{Notations:} 
Throughout the paper, we use uppercase letters to denote random variables, such as $X$, and the corresponding lowercase letters as their realizations, such as $x$. Whenever applicable, the corresponding calligraphic letters represent their alphabets, such as $\mathcal{X}$. For some set $\mathcal{X}$, denote by $|\mathcal{X}|$ its carnality.
The mutual information, entropy, and differential entropy are denoted by $\mathsf{I}(\cdot;\cdot)$, $\mathsf{H}(\cdot)$, and $\mathsf{h}(\cdot)$, respectively.
We denote by $\mathbb{P}[\cdot]$ the probability of a random event. For a random variable $X$, we denote its probability distribution, expectation, and variance by $P_X$, $\mathbb{E}[X]$, and $\mathbb{V}[X]$, respectively. If $X$ is uniformly distributed within an interval $[a,b]$, we write $X\sim{\rm Unif}([a,b])$, or if $X$ is a discrete uniform random variable, we write $X\sim{\rm Unif}(\mathcal{X})$. Specifically, for an integer $m>1$, if $\mathcal{X}$ has $m$ equiprobable mass points that are evenly-spaced within $[0,a]$, we write\footnote{ESDU stands for evenly spaced discrete uniform.} $X\sim{\rm ESDU}(a,m)$, which is equivalent to $X\sim{\rm Unif}\bigl(\mathcal{G}_{a,m}\bigr)$, where $\mathcal{G}_{a,m}=\bigl\{\frac{ia}{m-1}\bigr\}_{i=0}^{m-1}$. Besides, for an alphabet drawn from an evenly-spaced grid such as $\mathcal{G}_{a,m}$, we call it a $\frac{a}{m-1}$-spaced alphabet, where $\frac{a}{m-1}$ is the grid spacing.
We write $X\sim\mathcal{N}(\mu,\sigma^2)$ when $X$ is a Gaussian random variable with mean $\mu$ and variance $\sigma^2$.
For a random variable $X$, the notation $X^{(n)}$ denotes a sequence of $n$ independent and identically distributed instances of $X$. 
We denote by $\mathbb{R}$ and $\mathbb{N}$ the set of real numbers and natural numbers, respectively. For $x\in\mathbb{R}$, we define its quantization and quantization error under quantization intervals of width $a\in\mathbb{R}$ by $(x)_{a}^{\rm q}=x-x\bmod a$ and $(x)_{a}^{\rm e}=x\bmod a$, respectively, where $\rm{mod}$ represents the modulo operation. Finally, we define $[x]^+=\max\{0,x\}$ and by $\log(x)$ the base-2 logarithm of $x$.

\section{The Discrete-State Dirty Paper Channel: Motivation and Model} \label{sec:pre}
In this section, we first motivate the discrete-state DPC through OWC BC, then we define the class of discrete-state DPC that arises in the BC, as well as set the objective of this paper.

\subsection{The Peak-Constrained Broadcast Channel}
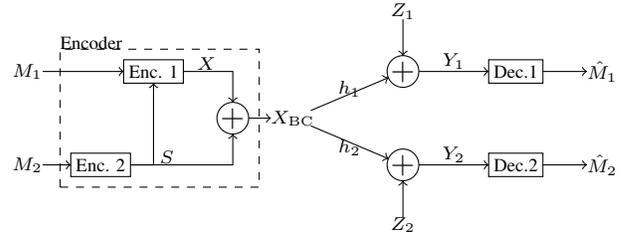
\begin{figure}
	\centering
	\begin{tikzpicture}[node distance = 1.5cm, inner sep=0, outer sep=0, font=\scriptsize] 
		\node (start) 	[] 												{};
		\node (m1) 		[above of=start, yshift=-25] 					{$M_1$};
		\node (m2) 		[below of=start, yshift=25] 					{$M_2$};
		\node (enc1) 	[draw, right of=m1, inner sep=2, xshift=5]  	{Enc. 1};
		\node (enc2) 	[draw, right of=m2, inner sep=2, xshift=-15]  	{Enc. 2};
		\node (osumEnc) [circle, draw,  right of =start, xshift=35] 	{\large $+$};
		\node (enc) 	[draw, inner sep=4, dashed, fit=(enc1) (enc2) (osumEnc)] {};
		\node at (enc.north west) [above right,node distance=0 and 0] {Encoder};
		\node (x)     	[right of=osumEnc, xshift=-20] 					{$X_{\rm BC}$};
		\node (temp)  	[right of=x, xshift=-1]							{};
		\node (osum1) 	[circle, draw,  above of =temp, yshift=-25] 	{\large $+$};
		\node (z1) 		[above of=osum1, yshift=-20]					{$Z_1$};
		\node (dec1)  	[draw, right of=osum1, inner sep=2] 			{ Dec.1};
		\node (end1)  	[right of=dec1, xshift=-10] 					{$\hat{M}_1$};
		\node (osum2) 	[circle, draw,  below of =temp, yshift=25] 		{\large $+$};
		\node (z2) 		[below of=osum2, yshift=20]						{$Z_2$};
		\node (dec2)  	[draw, right of=osum2, inner sep=2] 			{ Dec.2};
		\node (end2)  	[right of=dec2, xshift=-10] 					{$\hat{M}_2$};
		
		\draw[->] (m1) 			-- (enc1);
		\draw[->] (m2)			-- (enc2);
		\draw[->] (enc1.east) 	-| node[xshift=-10,yshift=3]{$X$} (osumEnc.north);
		\draw[->] (enc2.east) 	-| (osumEnc.south);
		\draw[->] (enc2.east) 	-| node[xshift= 5,yshift=3]{$S$} (enc1.south);
		\draw[->] (osumEnc)		-- (x);
		\draw[->] (x)			-- node[yshift=2]{$h_1$} (osum1) ;
		\draw[->] (z1)			-- (osum1) ;
		\draw[->] (osum1)		--node[anchor=south, yshift=1] {$Y_1$} (dec1) ;
		\draw[->] (dec1)		-- (end1) ;
		\draw[->] (x)			-- node[yshift=-2]{$h_2$} (osum2) ;
		\draw[->] (z2)			-- (osum2) ;
		\draw[->] (osum2)		--node[anchor=south, yshift=1] {$Y_2$} (dec2) ;
		\draw[->] (dec2)		-- (end2) ;
	\end{tikzpicture}
	\caption{The two-user peak-constrained Gaussian broadcast channel (BC) that adopts dirty paper coding.}
	\label{fig:bc}
\end{figure}

Consider a two-user broadcast channel (BC) under a peak-amplitude constraint as depicted in Fig. \ref{fig:bc} or expressed below:
\begin{equation}\label{eq:bc}
	Y_i = h_i X_{\rm BC} + Z_i, \; i=1,2
\end{equation}
where the transmitter wants to send private messages $M_1$ to user 1 and $M_2$ to user 2 by transmitting {\color{red}$X_{\rm BC}\in[0,\mathsf{P}]$} through a noisy link with channel gain $h_i$ and Gaussian noise $Z_i\sim\mathcal{N}(0,\sigma_i^2)$, $i=1,2$. Without loss of generality, we let $h_1=h_2=1$, since their difference can be reflected by the noise variances. We also assume that $\sigma_1<\sigma_2$ in this paper, without loss of generality. 

If $M_1$ and $M_2$ are encoded independently, both users will suffer IUI. To combat IUI at one of the users, such as user 1, the transmitter encodes $M_2$ to obtain a subcodeword, denoted by $S^{(n)}$, where $n$ is the code length. Then, the transmitter encodes $(M_1,S^{(n)})$ to obtain another subcodeword $X^{(n)}$, and finally transmits $X_{\rm BC}^{(n)}=X^{(n)}+S^{(n)}$. In such a way, $S^{(n)}$ contributes in generating $X^{(n)}$. Since we only consider memoryless channels, $n$ will be omitted, henceforth. Intuitively, this encoding process is similar to dirty paper coding \cite{Costa1983,yu2001trellis}. Through this encoding process, the BC in \eqref{eq:bc} can be seen equivalently as two channels as follows: a P2P channel with input $S$ and output $Y_2=S+X+Z_2$, where $X$ is interference, and a dirty paper channel (DPC) with input $X$ and output $Y_1=X+S+Z_1$, where $S$ is the known channel state at the encoder.

Through the DPC-based broadcasting scheme, a rate pair $(\mathtt{R}_1,\mathtt{R}_2)$ is achievable for the BC, such that 
\begin{subequations}
	\label{eq:bc_dpc_rate}
	\begin{align}
		\mathtt{R}_1 &\le \mathsf{I}(T;Y_1) - \mathsf{I}(T;S), \label{eq:bc_dpc_rate_r1} \\
		\mathtt{R}_2 &\le \mathsf{I}(S;Y_2), \label{eq:bc_dpc_rate_r2}
	\end{align}
\end{subequations}
where \eqref{eq:bc_dpc_rate_r1} is obtained from Gelfand-Pinsker (GP) Theorem \cite[Theorem 7.3]{book-NIT} under a given $P_T$ and mapping $x(t,s)$.
Note that, if $X$ and $S$ are independent, we have $\mathsf{I}(S;Y_2)= \mathsf{I}(X+S; Y_2) - \mathsf{I}(X;X+Z_2)$ in \eqref{eq:bc_dpc_rate_r2}, which will be helpful to generate the numerical results in Sec. \ref{sec:figures}. 

\subsection{The Discrete Dirty Paper Channel}
We want to investigate a discrete signaling scheme for the BC in \eqref{eq:bc} so that $S$ in \eqref{eq:bc_dpc_rate} is discretely distributed. Specifically, when $S$ is drawn from an evenly-spaced discrete grid, the following discrete-state DPC arises.
\begin{definition}\label{def:dpc_spec}
	The class of peak-constrained discrete-state DPC studied in this paper is defined as follows:
	\begin{equation} \label{eq:dpc_spec}
		Y = X+S+Z
	\end{equation}
	where $Y$ is the channel output, $X$ is the channel input that is subject to a peak constraint $X\in[0,\mathsf{A}]$, $Z\sim\mathcal{N}(0,\sigma^2)$ is Gaussian noise, and $S$ is the discrete channel state with alphabet $\mathcal{S}\subseteq\mathcal{G}_{\mathsf{B},\mathsf{N}}$, for some $\mathsf{B}>0$ and integer $\mathsf{N}>2$. 
\end{definition}

Denote the capacity of the DPC in Definition \ref{def:dpc_spec} by $\mathsf{C}$. Theorem \ref{thm:UB} below states an upper bound for $\mathsf{C}$. 
\begin{thm}\label{thm:UB}
	The capacity of the DPC in \eqref{eq:dpc_spec} satisfies that $\mathsf{C}\le \overline{\mathsf{C}}(\mathsf{A},\sigma)$, where
	\begin{equation} \label{eq:dpc_ub}
		\overline{\mathsf{C}}(\mathsf{A},\sigma) \triangleq \min\Biggl\{\frac{1}{2}\log\Bigl(1+\frac{\mathsf{A}^2}{4\sigma^2}\Bigr),\log\Bigl(1+\frac{\mathsf{A}}{\sqrt{2\pi e}\sigma}\Bigr)\Biggr\}
	\end{equation}
\end{thm}
\begin{proof} The proof is based on the GP theorem \cite[Theorem 7.3]{book-NIT} and can be found in Appendix \ref{app:UB}.
\end{proof}

This rest of this paper focuses on studying the ultimate achievable rate of the DPC in \eqref{eq:dpc_spec}, through existing or new coding/precoding schemes. Next, a benchmark achievable rate is obtained via Costa's coding.

\section{Benchmark Achievable Rate of the Discrete DPC} \label{sec:r0}
In this section, we apply Costa's coding and Tomlinson-Harashima (TH) precoding to the DPC in \eqref{eq:dpc_spec}, and obtain a benchmark achievable rate for it. Costa's coding \cite{Costa1983} is designed for the power-constrained Gaussian DPC. The concept is to construct the codeword through random binning followed by partially presubtracting the state. In Costa's coding, $T$ in the GP Theorem \cite[Theorem 7.3]{book-NIT} is chosen to be $T=X+\alpha S$, wherein $X$ and $S$ are independent and $\alpha$ is optimally chosen to maximize the achievable rate. While this construction achieves the capacity of the power-constrained Gaussian DPC, its achievable rate in the peak-constrained DPC in \eqref{eq:dpc_spec} has not been studied. Next, a lower bound on the achievable rate of Costa's coding in the DPC in \eqref{eq:dpc_spec} is stated in Theorem \ref{thm:costa}, which serves as a benchmark for comparison.
\begin{thm} \label{thm:costa}
	For the DPC in \eqref{eq:dpc_spec}, we have $\mathsf{C}\ge \mathsf{R}_0$, where
	\begin{align}
		\mathsf{R}_0 &\triangleq \frac{1}{2}\log\Bigl(\frac{12}{2\pi e} + \frac{\mathsf{A}^2}{2\pi e\sigma^2}\Bigr). \label{eq:r0}
	\end{align} 
\end{thm}
\begin{proof} The proof can be found in Appendix \ref{app:R0}, where a construction similar to the one in Costa's coding is adopted.
\end{proof}

A structured algorithm to realize Costa's coding is not available. Fortunately, however, Tomlinson-Harashima (TH) precoding proposed in \cite{tomlinson1971new,harashima1972matched} can be used in \cite{yu2005trellis,erez2005capacity}. It can be shown that TH precoding achieves $\mathsf{R}_0$ in the DPC in \eqref{eq:dpc_spec}, providing a structured algorithm. The proof is straightforward following the proof of \cite[Lemma 6]{erez2005capacity} and is hence omitted here. Next, we propose new precoding schemes for the peak-constrained DPC in \eqref{eq:dpc_spec}, which will be compared with the benchmark in \eqref{eq:r0} in Sec. \ref{sec:figures}.

\section{Proposed Precoding Schemes} \label{sec:newR}
To achieve a larger rate than $\mathsf{R}_0$, we propose two new precoding schemes for the DPC in \eqref{eq:dpc_spec} in this section, as described next. Both proposed scheme will be shown to outperform $\mathsf{R}_0$ in Sec. \ref{sec:figures}. 
\subsection{Proposed Precoding Scheme 1}
Scheme 1 can be considered as a variant of TH precoding (c.f. \cite[Lemma 6]{erez2005capacity}) which results to an equivalent modulo channel, but the encoding process in Scheme 1 is based on a discrete alphabet where the state will be quantized, which is different from the TH precoding. In details, the operations at the transmitter and the receiver of Scheme 1 are given as follows:
\begin{enumerate} 
	\item Transmitter: Let $T\sim{\rm ESDU}(\mathsf{A},\mathsf{K})$ be used for constructing the codeword of $M$. Let $\Delta=\frac{\mathsf{A}}{\mathsf{K}-1}$ and denote $\mathsf{A}_{\Delta}=\mathsf{A}+\Delta$. To send $M$ given $S$, transmit 
	\begin{equation}
		x = \bigl[t - (\alpha s+w)_{\Delta}^{\rm q} - w'\bigr] \bmod \mathsf{A}_\Delta,
	\end{equation}
	where $W\sim{\rm Unif}([0,\mathsf{A}_\Delta])$ and $W'\sim{\rm ESDU}(\mathsf{A},\mathsf{K})$ are the common random variables whose realizations are known at both the transmitter and the receiver;
	\item Receiver: Upon receiving $Y$, obtain $Y'$ as follows. 
	\begin{subequations}
		\begin{align}
			y' &= \bigl[\alpha y + w + w'\bigr] \bmod \mathsf{A}_\Delta \\
			& = \bigl[t-t+\alpha (x+s+z) + w + w'\bigr] \bmod \mathsf{A}_\Delta \\
			& = \bigl[ t + (\alpha s+w)_{\Delta}^{\rm e} + (\alpha-1)x + \alpha z \bigr] \bmod \mathsf{A}_\Delta  \\
			& = \bigl[ t + w_0 + (\alpha-1)x + \alpha z \bigr] \bmod \mathsf{A}_\Delta  \\
			& = \bigl[ t + \tilde{z}_\alpha' \bigr] \bmod \mathsf{A}_\Delta,
		\end{align}
	\end{subequations}
	where $w_0\triangleq[\alpha s+w]\bmod \Delta$ and $\tilde{z}_\alpha'\triangleq w_0 + (\alpha-1)x + \alpha z$. 
\end{enumerate}
\begin{remark} We have the following remarks for Scheme 1.
	\begin{enumerate}
		\item Since the modulo is over $\mathsf{A}_\Delta=\mathsf{K}\Delta$ on a $\Delta$-spaced alphabet, the input peak constraints still holds after precoding, i.e., $X\in[0,\mathsf{A}]$;
		\item Since $W\bmod\Delta\sim{\rm Unif}([0,\Delta])$, we have $W_0\sim{\rm Unif}\bigl([0,\Delta]\bigr)$ and $W_0$ is independent of $T$, $S$, $X$, and $Z$, according to the Crypto Lemma \cite[Lemma 1]{erez2004achieving}; 
		\item $T$ is independent of $\tilde{Z}_\alpha'$, which follows 2) above;
		\item Scheme 1 forms an equivalent channel that has input $T$ and output $[T+\tilde{Z}_\alpha']\bmod \mathsf{A}_\Delta$, where $\tilde{Z}_\alpha'$ is the effective modulo-channel noise. 
	\end{enumerate}
\end{remark}
The achievable rate of Scheme 1 can be lower bound by $\mathsf{R}_1$ as given in Theorem \ref{thm:r1}.
\begin{thm}\label{thm:r1}
	For the DPC in \eqref{eq:dpc_spec}, we have $\mathsf{C}\ge \mathsf{R}_1$, where
	\begin{multline} \label{eq:r1}
		\mathsf{R}_1\triangleq \frac{1}{2}\log\Bigl(\mathsf{A}^2 + \frac{2\mathsf{A}^2}{\mathsf{K}-1} + 12\sigma^2\Bigr) - \frac{1}{2}\log(2\pi e)  \\
		- \frac{1}{2}\log\Bigl( \frac{\mathsf{A}^2}{12(\mathsf{K}-1)^2}\Bigl[1-\frac{1}{(\mathsf{K}-2)^2}\Bigr]+\sigma^2 \Bigr)
	\end{multline}
	achieved via Scheme 1.
\end{thm}
\begin{proof} The proof can be found in Appendix \ref{app:thm_R1}. 
\end{proof}

Note that, with a proper choice of $\mathsf{K}$, $\mathsf{R}_1$ is guaranteed to outperform $\mathsf{R}_0$. The proof is straightforward by comparing $\mathsf{R}_0$ and $\mathsf{R}_1$ and is omitted here. This conclusion will also be observed later in Sec. \ref{sec:figures}, where the achievable rates of different precoding schemes are compared numerically.

\subsection{Proposed Precoding Scheme 2}
Scheme 2 is based on a mapping function $\varphi:\mathcal{T}\times\mathcal{S}\mapsto\mathcal{U}$, where $\mathcal{T},\mathcal{S},\mathcal{U}\subseteq\{j\delta|j\in\mathbb{N}\}$, defined as follows:
\begin{equation}\label{eq:varphi}
	u=\varphi(t,s)\triangleq\bigl( t + (|\mathcal{T}|-1)s \bigr) \bmod \bigl(|\mathcal{T}|\delta\bigr) + s.
\end{equation}
An example is provided to help explain the behavior of the mapping $\varphi$, as follows.

\emph{Example:} Let $\delta=1$, $\mathcal{T}=\{0,1,2\}$, and $\mathcal{S}=\{0,1,\dots,4\}$. Then, Table \ref{tab:phi} shows the result of $\varphi(t,s)$.
\begin{table} [!h]
	\centering
	\caption{}
	\label{tab:phi}
	\begin{tabular}{ c c | c c c c c}
		\multicolumn{2}{c|}{\multirow{2}{*}{$\varphi(t,s)$}}& \multicolumn{5}{c}{$s$} \\ \cline{3-7}
		&& 0 & 1 & 2 & 3 & 4\\\hline
		\multirow{3}{*}{$t$}&\multicolumn{1}{|c|}{0} & 0 & 3 & 3 & 3 & 6\\
		&\multicolumn{1}{|c|}{1} & 1 & 1 & 4 & 4 & 4\\
		&\multicolumn{1}{|c|}{2} & 2 & 2 & 2 & 5 & 5
	\end{tabular}
\end{table} 

The operations at the transmitter and the receiver for Scheme 2 are given as follows:
\begin{enumerate}
	\item \emph{Transmitter:} Define $\delta=\frac{\mathsf{B}}{\mathsf{K}-1}$, $\mathsf{K}'=\lfloor \frac{\mathsf{A}}{\delta} \rfloor+1$, and $\mathsf{A}'=(\mathsf{K}'-1)\delta$. Let $T\sim{\rm ESDU}(\mathsf{A}',\mathsf{K}')$ be used for constructing the codeword of $M$. To send $M$ given $S$, obtain $u=\varphi(t,s)=\bigl[ t + (\mathsf{K}'-1)s \bigr] \bmod \mathsf{A}_\delta'+s$ through the mapping function $\varphi$ first, where $\mathsf{A}_\delta'=\mathsf{A}'+\delta$, then transmit 
	\begin{equation}
		x=u-s.
	\end{equation}
 	\item \emph{Receiver:} Upon receiving $Y$, the receiver decodes $T$, where we note that the following holds
 		\begin{equation}
 			y = x+s+z
 			 = u + z.
 		\end{equation}
\end{enumerate}
\begin{remark} We have the following remarks for Scheme 2.
	\begin{enumerate}
		\item The mapping $\mathcal{U}\mapsto\mathcal{T}$ is a surjective mapping (as exemplified in Table \ref{tab:phi}), where $\mathcal{U}$ is the alphabet of $U=\varphi(T,S)$. Thus, if $U$ can be detected successfully from $Y=U+Z$,  then $T$ and the message $M$ can be obtained successively;
		\item The alphabets of the codeword and the state are all $\delta$-spaced, i.e., $\mathcal{T},\mathcal{S}\subseteq\{j\delta|j\in\mathbb{N}\}$;
		\item $X$ is independent of $S$, which can be seen from Table \ref{tab:phi} where, given $S\in\mathcal{S}$, the distribution of $X=\varphi(T,S)-S$ does not change;
		\item Scheme 2 does not necessarily make use of the whole range of feasible input intensity , i.e., $X\in[0,\mathsf{A}]$, since $\delta$ generally leads to $\mathsf{A}'\le\mathsf{A}$. However, in a DPC-based broadcasting scheme, $\delta$ can be designed to optimize Scheme 2. This will be explained with more details in the next section.
	\end{enumerate}
\end{remark}

Next, Theorem \ref{thm:r2} gives the achievable rate of Scheme 2, which is denoted by $\mathsf{R}_2$.

\begin{thm}\label{thm:r2}
	For the DPC in \eqref{eq:dpc_spec}, we have $\mathsf{C}\ge \mathsf{R}_2$, where
	\begin{subequations}
		\label{eq:r2}
		\begin{align} 
			\mathsf{R}_2 & \triangleq \mathsf{I}(T;Y) \\
						&=\mathsf{h}(U+Z) - \mathsf{h}(U+Z|T),
		\end{align}
	\end{subequations}
	where $U = \varphi(T,S)$, wherein $T\sim{\rm ESDU}(\mathsf{A}',\mathsf{K}')$, $\mathsf{A}'=(\mathsf{K}'-1)\delta$, and $\mathsf{K}'=\lfloor \frac{\mathsf{A}}{\delta} \rfloor+1$.
\end{thm}
\begin{proof}
	This achievable rate is obtained based on the GP Theorem \cite[Theorem 7.3]{book-NIT} by noting that the choice of auxiliary variable and mapping function in Scheme 2 is not optimal, and $T$ and $S$ are independent.
\end{proof}
Note that, $U$ in \eqref{eq:r2} is discrete and is independent of the Gaussian random variable $Z$, so that $\mathsf{h}(U+Z)$ and $\mathsf{h}(U+Z|T)$ can be approximated given the probability mass functions (p.m.f.-s), $P_U$ and $P_{U|T}$.

\section{Numerical Results} \label{sec:figures}
In this section, we first numerically compare the achievable rates of all the precoding schemes that has been discussed so far. Then, we show how Scheme 2 can help to enlarge the achievable rate region of the peak-constrained BC in \eqref{eq:bc}.
In all simulations, we let $\sigma=1$ for the DPC, and $\sigma_1=1$, $h_1=h_2=1$ for the BC, without loss of generality. In the discussion below, we define SNR$=10\log_{10}\bigl(\frac{\mathsf{A}}{\sigma}\bigr)$, SNR$_1=10\log_{10}\bigl(\frac{\mathsf{P}}{\sigma_1}\bigr)$, and INR$=10\log_{10}\bigl(\frac{\mathsf{B}}{\sigma}\bigr)$.

\subsection{Achievable Rates for the DPC in \eqref{eq:dpc_spec}}
In this simulation, we compare the achievable rates of all the precoding schemes, i.e., $\mathsf{R}_i$, $i=1,2,3$, in the DPC in \eqref{eq:dpc_spec}. Since we are interested in the application of Scheme 2 in the DPC-based broadcasting (demonstrated in the next subsection), thus, we will constrain the state $S$ to take values from the same grid as the channel input's in this simulation. We will also test the effect of different state strength through several reference INR values. 
In details, we first set a reference grid spacing $\Delta_0=3\sigma$, which helps to generate an evenly-spaced grid $\Delta\mathbb{N}$, where $\Delta=\frac{\mathsf{A}}{\mathsf{K}-1}$ and $\mathsf{K}=\min\bigl\{2, \lceil \frac{\mathsf{A}}{\Delta_0} \rceil \bigr\}$. Then, we let $T\sim{\rm ESDU}(\mathsf{A},\mathsf{K})$ so that $\mathcal{T}\subseteq\Delta\mathbb{N}$. To generate the state $S$, given a reference INR value, we obtain the corresponding $\mathsf{B}$ first, then we can choose an $\mathcal{S}\subseteq\Delta\mathbb{N}$ such that $S\in[0,\mathsf{B}+\Delta]$, and specify a distribution over $\mathcal{S}$ to generate $S$. More specifically, as an example in this simulation, we let $S\sim{\rm ESDU}\bigl(\mathsf{B}',\frac{\mathsf{B}'}{\Delta}+1\bigr)$, where $\mathsf{B}'=\lceil \frac{\mathsf{B}}{\Delta} \rceil\Delta$. Given $(\mathcal{T},\mathcal{S})$, we obtain $\mathsf{R}_2$ through \eqref{eq:r2}. As for $\mathsf{R}_0$ and $\mathsf{R}_1$ in this simulation, both of them only depends on SNR, where we maximize $\mathsf{R}_1$ over $\mathsf{K}$ in \eqref{eq:r1}. 

\begin{figure}
	\centering
%
%
%
\definecolor{mycolor1}{rgb}{0.00000,0.44700,0.74100}%
\definecolor{mycolor2}{rgb}{0.85000,0.32500,0.09800}%
\definecolor{mycolor3}{rgb}{0.92900,0.69400,0.12500}%
\definecolor{mycolor4}{rgb}{0.49400,0.18400,0.55600}%
\definecolor{mycolor5}{rgb}{0.46600,0.67400,0.18800}%

\begin{tikzpicture}
	
	\begin{axis}[%
		width=.95\columnwidth,
		height=.7\columnwidth,
		xmin=0,
		xmax=20,
		xlabel={\scriptsize SNR (dB)},
		ymin=-0,
		ymax=5,
		ytick={0,1,2,3,4,5},
		ylabel={\scriptsize Rate (bits/Tr.)},
		xmajorgrids,
		ymajorgrids,
		ticklabel style = {font=\scriptsize},
		xlabel near ticks,
		ylabel near ticks,
		legend style={ at={(axis cs: 0, 5)}, anchor=north west}
		]
		\addplot [color=mycolor1]
		table[row sep=crcr]{%
			0	0.160964047443681\\
			1	0.240764845172548\\
			2	0.351537769479087\\
			3	0.498291242926011\\
			4	0.682956368898559\\
			5	0.819814175671874\\
			6	0.973282700318478\\
			7	1.14582506250897\\
			8	1.33727268473667\\
			9	1.54697680854408\\
			10	1.77387282394493\\
			11	2.01657866458013\\
			12	2.27350918813765\\
			13	2.54298866830575\\
			14	2.82334783530117\\
			15	3.11299800861738\\
			16	3.41048059767537\\
			17	3.71449433269512\\
			18	4.02390473640102\\
			19	4.33774090510789\\
			20	4.65518421732466\\
		};
		\addlegendentry{\scriptsize $\overline{\mathsf{C}}$ \eqref{eq:dpc_ub}}
		
		
		\addplot [color=black]
		table[row sep=crcr]{%
			0	-0.196875726110095\\
			1	-0.165129926333992\\
			2	-0.117514002414838\\
			3	-0.0479494577461771\\
			4	0.0501675529088417\\
			5	0.182620224138008\\
			6	0.352679461115438\\
			7	0.559944747559846\\
			8	0.800493737506549\\
			9	1.0682339436597\\
			10	1.35658187584816\\
			11	1.6596731900255\\
			12	1.97288112805923\\
			13	2.29283299411126\\
			14	2.61719402393576\\
			15	2.94440120216651\\
			16	3.27343046810866\\
			17	3.60362005423136\\
			18	3.93454604684122\\
			19	4.26593840268758\\
			20	4.59762570266352\\
		};
		\addlegendentry{\scriptsize $\mathsf{R}_0$ \eqref{eq:r0} }

		\addplot [color=black, dashed, line width=.5pt]
		table[row sep=crcr]{%
			0	-0.137381708001552\\
			1	-0.0819818232485347\\
			2	-0.00826657218553072\\
			3	0.0833958778359568\\
			4	0.195934228831417\\
			5	0.330348165901093\\
			6	0.496256519731266\\
			7	0.686721750913805\\
			8	0.907323165399429\\
			9	1.15100421130215\\
			10	1.41782159236716\\
			11	1.70297071012594\\
			12	2.00245090638984\\
			13	2.3125152491957\\
			14	2.63006014912296\\
			15	2.95270950989164\\
			16	3.27875144664791\\
			17	3.6070095374714\\
			18	3.93669769828457\\
			19	4.26730124644347\\
			20	4.59848770049582\\
		};
		\addlegendentry{\scriptsize $\mathsf{R}_1$ \eqref{eq:r1}}

		\addplot [only marks, mark=x, mark size = 2pt, mark options={solid, mycolor3, line width=.5pt}]
		table[row sep=crcr]{%
			0	0.021475737806842\\
			1	0.0573409627254922\\
			2	0.185649907088862\\
			3	0.323545804928075\\
			4	0.509961670631976\\
			5	0.792911452667763\\
			6	0.910932652831019\\
			7	0.975581883071143\\
			8	1.25138238038816\\
			9	1.46493428565391\\
			10	1.73598623572931\\
			11	1.95237792065529\\
			12	2.24218109408084\\
			13	2.5032990693514\\
			14	2.77522587717658\\
			15	3.06782224721316\\
			16	3.37282666663398\\
			17	3.67666122849553\\
			18	3.99426526755333\\
			19	4.30804416478921\\
			20	4.62837703032135\\
		};
		\addlegendentry{\scriptsize $\mathsf{R}_2$ \eqref{eq:r2} (ref. INR=5dB)}

		\addplot [only marks, mark=o, mark size = 2pt, mark options={solid, mycolor1, line width=.5pt}]
		table[row sep=crcr]{%
			0	0.00849792532700633\\
			1	0.02366798863105\\
			2	0.0701558658039998\\
			3	0.189477819814398\\
			4	0.417443197598985\\
			5	0.651891681856035\\
			6	0.857468310517578\\
			7	0.965677427259436\\
			8	1.2137429691557\\
			9	1.43111363390039\\
			10	1.68662830819022\\
			11	1.93141012585206\\
			12	2.19769101546687\\
			13	2.47037577722275\\
			14	2.76349634591076\\
			15	3.05842305751913\\
			16	3.3647312085216\\
			17	3.67040861211785\\
			18	3.98883834790593\\
			19	4.30381882629043\\
			20	4.62494563364028\\
		};
		\addlegendentry{\scriptsize $\mathsf{R}_2$ \eqref{eq:r2} (ref. INR=10dB)}
		
		\addplot [only marks, mark=square, mark size = 2pt, mark options={solid, mycolor5, line width=.5pt}]
		table[row sep=crcr]{%
			0	0.000916952975186014\\
			1	0.00493342922985462\\
			2	0.0331795835453885\\
			3	0.133159948939873\\
			4	0.335996773699454\\
			5	0.599706907641695\\
			6	0.826231641257424\\
			7	0.952617229730044\\
			8	1.17358005797442\\
			9	1.40794242479565\\
			10	1.65208611228843\\
			11	1.90575821359901\\
			12	2.17687857080668\\
			13	2.4551774792118\\
			14	2.74939898194283\\
			15	3.04719960270359\\
			16	3.35506796150305\\
			17	3.66303965592995\\
			18	3.98240589227891\\
			19	4.29887434261162\\
			20	4.62095354165522\\
		};
		\addlegendentry{\scriptsize $\mathsf{R}_2$ \eqref{eq:r2} (ref. INR=20dB)}

	\end{axis}
	
\end{tikzpicture}%
	\caption{Compare the achievable rate of Costa's coding and the proposed precoding schemes in the DPC in \eqref{eq:dpc_spec}.}
	\label{fig:spec}
\end{figure}
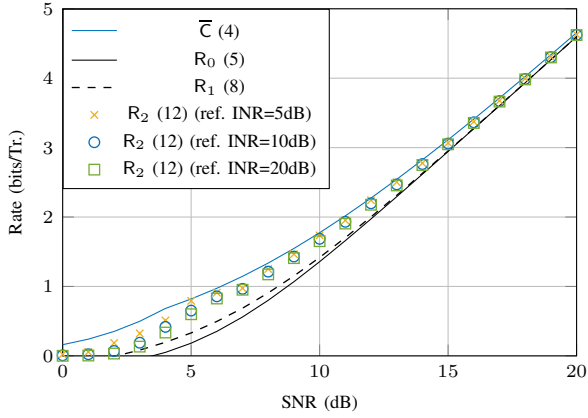

Fig. \ref{fig:spec} compares $\mathsf{R}_0$ \eqref{eq:r0}, $\mathsf{R}_1$ \eqref{eq:r1}, and $\mathsf{R}_2$ \eqref{eq:r2}, where $\mathsf{R}_1$ is obtained by maximizing \eqref{eq:r1} over $\mathsf{K}$. It can be seen that, Scheme 2 outperforms both Scheme 1 and TH precoding under all tested INR and SNR, where TH precoding performs the worst among all settings. Besides, when SNR$\ge 6$dB, Scheme 2 achieves a rate that is close to the capacity under all tested INR, which varies from $5$ to $20$dB, and when SNR$<6$dB, $\mathsf{R}_2$ decrease when INR increase. Notably, at high SNR, all precoding schemes in the simulation tends to approximate the capacity upper bound $\overline{\mathsf{C}}(\mathsf{A},\sigma)$, which characterizes the capacity of the DPC in \eqref{eq:dpc_spec} at high SNR. Also, the close-to-capacity performance when SNR$\ge6$dB shows the promising efficiency of Scheme 2 when adopted in the DPC-based broadcasting, as shown in the next simulation. 

\subsection{New Inner Bound of the Two-User Peak-Constrained Gaussian BC}
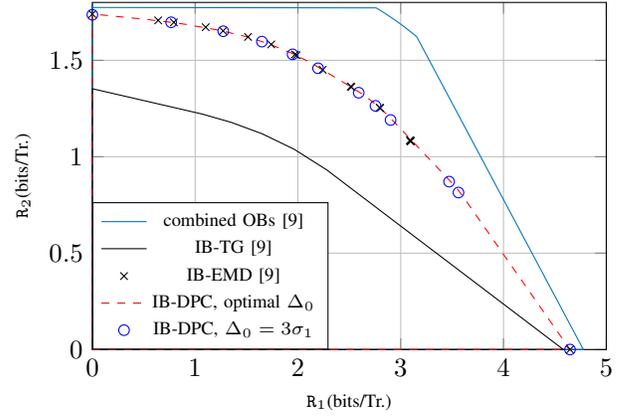
\begin{figure}
	\centering
%
%
\definecolor{mycolor1}{rgb}{0.00000,0.44700,0.74100}%
\definecolor{mycolor2}{rgb}{1.00000,0.00000,1.00000}%
\begin{tikzpicture}

\begin{axis}[%
width=.95\columnwidth,
height=.7\columnwidth,
xmin=0,
xmax=5,
xlabel={\scriptsize $\mathtt{R}_1$(bits/Tr.)},
ymin=0,
ymax=1.8,
ytick={0,.5, 1, 1.5, 2},
ylabel={\scriptsize $\mathtt{R}_2$(bits/Tr.)},
xmajorgrids,
ymajorgrids,
ylabel near ticks,
xlabel near ticks,
legend style={at={(axis cs: 0,0)}, anchor=south west}
]

\addplot [color=mycolor1]
table[row sep=crcr]{%
	0		2.039652629	\\
	0		0	\\
	4.777117572		0	\\
	3.156254359		1.621773925	\\
	3.148183849		1.62535669	\\
	3.140068526		1.628936887	\\
	3.131907896		1.63251439	\\
	3.123701459		1.636089069	\\
	3.115448704		1.639660793	\\
	3.107149113		1.643229431	\\
	3.098802158		1.646794847	\\
	3.090407305		1.650356907	\\
	3.081964007		1.653915472	\\
	3.073471712		1.657470403	\\
	3.064929855		1.661021559	\\
	3.056337864		1.664568797	\\
	3.047695156		1.668111972	\\
	3.039001138		1.671650939	\\
	3.030255207		1.675185548	\\
	3.021456749		1.67871565	\\
	3.012605139		1.682241094	\\
	3.003699742		1.685761724	\\
	2.994739912		1.689277387	\\
	2.98572499		1.692787924	\\
	2.976654306		1.696293176	\\
	2.967527178		1.699792983	\\
	2.958342911		1.703287182	\\
	2.949100799		1.706775608	\\
	2.939800122		1.710258094	\\
	2.930440145		1.713734472	\\
	2.921020123		1.717204572	\\
	2.911539294		1.720668221	\\
	2.901996884		1.724125246	\\
	2.892392103		1.727575471	\\
	2.882724145		1.731018718	\\
	2.872992192		1.734454806	\\
	2.863195408		1.737883556	\\
	2.85333294		1.741304783	\\
	2.843403921		1.744718302	\\
	2.833407466		1.748123925	\\
	2.823342672		1.751521465	\\
	2.813208618		1.754910729	\\
	2.803004367		1.758291525	\\
	2.79272896		1.761663658	\\
	2.782381422		1.765026932	\\
	2.771960755		1.768381148	\\
	2.761465945		1.771726106	\\
	0	1.77387282394493\\
};
\addlegendentry{\scriptsize combined OBs \cite{chaaban2016capacity} }

\addplot [color=black]
  table[row sep=crcr]{%
	0		0	\\
	4.582802923		0	\\
	2.280938576		0.931397313	\\
	1.962685757		1.038858408	\\
	1.652048925		1.118877022	\\
	1.352752965		1.177801745	\\
	1.069807162		1.221085576	\\
	0		1.352626118	\\
	0		1.352593273	\\
	0		1.352551922	\\
	0		1.352499863	\\
	0		1.352434322	\\
	0		1.352351808	\\
	0		1.352247922	\\
	0		1.352117129	\\
	0		1.351952456	\\
	0		1.351745121	\\
	0		1.351484065	\\
	0		1.351155358	\\
	0		1.350741447	\\
	0		1.350220219	\\
	0		1.349563799	\\
	0		1.348737049	\\
	0		1.34769565	\\
	0		0	\\
};
\addlegendentry{\scriptsize IB-TG \cite{chaaban2016capacity} }

\addplot [only marks, mark=x, color=black]
table[row sep=crcr]{%
	4.64706871941261	0\\
	3.09677109185915	1.07898721900429\\
	3.09585681804178	1.08151106275395\\
	3.09330149265034	1.08368542516442\\
	3.09020158932742	1.08557816529741\\
	2.80027943098631	1.2522273346399\\
	2.79543771831395	1.25431162320581\\
	2.52030068088184	1.36145933882316\\
	2.51365091050567	1.36377613106186\\
	2.24063365773261	1.4509219667361\\
	1.98717935003286	1.5262726326257\\
	1.97083661562541	1.53008088442827\\
	1.74015829690698	1.58225625977\\
	1.51464397713933	1.62126883810737\\
	1.27217540375245	1.6526816301787\\
	1.10273337740818	1.67163923074285\\
	0.792911139532705	1.69683771006929\\
	0.640509224626932	1.70690499006441\\
	0	1.74015885355563\\
};
\addlegendentry{\scriptsize IB-EMD \cite{chaaban2016capacity} }

\addplot [color=red, dashed]
table[row sep=crcr]{%
	0	0\\
	1.54722394727107	0\\
	1.55893079011886	0\\
	1.56987446299862	0\\
	1.58209376651152	0\\
	4.64706871941261	0\\
	3.60981518957172	0.790939911279723\\
	3.5504781155806	0.832836630204098\\
	3.4883921617912	0.869865796455496\\
	2.95758745822429	1.16790934402846\\
	2.86318401992552	1.22053124689358\\
	2.76195263297078	1.26909873077922\\
	2.65310645435079	1.31518852031473\\
	2.53541225436206	1.35598880223422\\
	2.11124570515246	1.491410910242\\
	1.93707280643468	1.53746676965392\\
	1.73930995739721	1.58074036795711\\
	1.6495152342147	1.59683661678812\\
	1.51075695278198	1.62085295864382\\
	1.24055095972144	1.65515851232364\\
	0.91171694736066	1.68604700838587\\
	0.766386773029875	1.69705553399503\\
	0.499124814860274	1.713531450325\\
	0	1.74\\
	0	1.58209376651152\\
	0	1.56987446299862\\
	0	1.55893079011886\\
	0	1.54722394727107\\
	0	0\\
};
\addlegendentry{\scriptsize IB-DPC, optimal $\Delta_0$ }

\addplot [only marks, mark=o, color=blue]
table[row sep=crcr]{%
	4.64706871941261	0\\
	3.56186084096565	0.814277182852624\\
	3.4707000341568	0.870770558189975\\
	2.90131882971589	1.19034864017992\\
	2.7545116023555	1.26374221995289\\
	2.59107946444211	1.3318646161587\\
	2.19552055459047	1.45857470763393\\
	1.94806902119302	1.53074311021037\\
	1.64951523421469	1.59683661678803\\
	1.27343491478426	1.65069562738323\\
	0.766386773029863	1.69705553399498\\
	0	1.73598623572929\\
};
\addlegendentry{\scriptsize IB-DPC, $\Delta_0=3\sigma_1$ }

 \end{axis}

\end{tikzpicture}%
	\caption{Inner and outer bounds of the two-user peak-constrained Gaussian BC under $\text{SNR}_1=20$ dB and $\sigma_2=10\sigma_1$.}
	\label{fig:bc_rate}
\end{figure}

For the two-user BC in \eqref{eq:bc}, we adopt Scheme 2, the DPC-based transmission scheme, to obtain a new inner bound (IB) for the BC. Two baseline IBs obtained in \cite{chaaban2016capacity} are compared with the new inner bound, one beseline IB is based on a truncated Gaussian (TG) distribution and have been shown to be tight at high SNR$_1$, and the other baseline IB is based on an entropy-maximizing discrete (EMD) distribution and superposition coding, where we set the alphabet size $K_1,K_2\le 40$ in \cite[Theorem 3]{chaaban2016capacity}. We also adopts the outer bound (OB) from \cite{chaaban2016capacity}, which is combined with two other straightforward OBs: $\mathtt{R}_i\le\overline{\mathsf{C}}(\mathsf{P},\sigma_i)$ and $\mathtt{R}_1+\mathtt{R}_2\le\overline{\mathsf{C}}(\mathsf{P},\sigma_1)$.
In this simulation, we are interested in a moderate SNR$_1$, such as 20 dB, which is more practical and the computation complexity to obtain the IB of Scheme 2 is not too high. 
In the simulation, we adopts the reference grid spacing, $\Delta_0$, to generate an evenly-spaced alphabet for constructing the codes, and we vary $\Delta_0$ within $[0.5\sigma_1,4\sigma_1]$ with a step size of $0.5\sigma_1$ to cover more settings. In details, for each $\Delta_0$, we first obtain $\mathsf{K}=\min\bigl\{2, \lceil \frac{\mathsf{P}}{\Delta_0} \rceil\bigr\}$, then we go through all possible $\mathsf{K}_1$ within the range $[1,\mathsf{K}]$. Let $\Delta=\frac{\mathsf{P}}{\mathsf{K}-1}$. For each $\mathsf{K}_1$, let $T\sim{\rm ESDU}\bigl((\mathsf{K}_1-1)\Delta,\mathsf{K}_1\bigr)$ and construct $\mathcal{S}\subseteq\mathcal{G}_{\mathsf{P}-(\mathsf{K}_1-1)\Delta,\mathsf{K}-\mathsf{K}_1+1}$, which results in $X+S\le\mathsf{P}$, and let $S\sim{\rm Unif}(\mathcal{S})$. Note that, each $(\mathsf{K},\mathsf{K}_1,\mathcal{S})$ setting corresponds to one set of p.m.f.-s of $T$, $W$, $S$, and $U$. Then, a achievable rate pair, $(\mathtt{R}_1,\mathtt{R}_2)$, can be numerically obtained through \eqref{eq:bc_dpc_rate}. Finally, the Scheme 2 based IB is the convex hull of the rate pairs obtained under all settings of $(\mathsf{K},\mathsf{K}_1,\mathcal{S})$, as described above. 
As shown in Fig. \ref{fig:bc_rate}, it can be seen that the new IB of Scheme 2 is much larger than the IB of a TG distribution under the tested setting and performs comparable to the second baseline IB, where $\text{SNR}_1=20$dB and $\sigma_2=10\sigma_1$. It also shows that $\Delta_0=3\sigma_1$ leads to a good approximation for the new inner bound under the tested setting.

\section{Conclusions} \label{sec:concl}
In summary, we studied a class of peak-constrained discrete-state DPC that arises in OWC broadcasting with discrete signaling, and proposed two new precoding schemes to study its capacity. The proposed schemes are shown to outperform a baseline performance achieved by Costa's coding and TH precoding. Notably, one of the proposed scheme is shown to achieve rates that are close to capacity. Then, for a two-user OWC BC, a new inner bound is obtained based on the developed precoding, which is shown to be comparable to the best known inner bound that is based on discrete distributions.

As for future work, applying Scheme 2 in a peak-constrained Gaussian BC with a vector input is our next step. Besides, the capacity of the peak-constrained DPC under a general-state case is open, where the proposed scheme 1 can be a baseline. Since this paper has shown that discrete signaling outperforms TG signaling, it will be interesting to design and study practical discrete signaling schemes for OWC BC that outperforms the well-accepted optical OFDM schemes, which generates TG distributed signals.

\appendices

\section{Proof of Theorem \ref{thm:UB}} \label{app:UB}
\begin{proof}
	For some $\epsilon_n>0$ such that $\lim_{n\to\infty}\epsilon_n=0$, using Fano's inequality, we have
	\begin{subequations}
		\begin{align}
			n\mathsf{C}=\mathsf{I}(M;Y^{(n)}) + n\epsilon_n &\le \mathsf{I}(M;Y^{(n)},S^{(n)}) + n\epsilon_n\\
			& = \mathsf{I}(M;Y^{(n)}|S^{(n)}) + n\epsilon_n\\
			& \le \mathsf{I}(M;X^{(n)}+Z^{(n)}) + n\epsilon_n\\
			& \le n\mathsf{C}_0(\mathsf{A},\sigma) + n\epsilon_n,
		\end{align}
	\end{subequations}
	where $\mathsf{C}_0(\mathsf{A},\sigma)$ denotes the capacity of a channel with input $X\in[0,\mathsf{A}]$ and output $X+Z$ where $Z\sim\mathcal{N}(0,\sigma^2)$, which is also the state-free capacity of the DPC in Definition \ref{def:dpc_spec}. Next, we have that $\mathsf{C}_0(\mathsf{A},\sigma)$ can be upper bounded by $\overline{\mathsf{C}}(\mathsf{A},\sigma)$ (defined in \eqref{eq:dpc_ub}) as shown in \cite[eq. (19)]{Lapidoth-Moser-Wigger-2009} and \cite[eq. (1)]{mckellips2004simple}, resulting in $\mathsf{C}\le\overline{\mathsf{C}}(\mathsf{A},\sigma)$. This ends the proof.
\end{proof}

\section{Proof of Theorem \ref{thm:costa}} \label{app:R0}
\begin{proof}
In the capacity expression in the GP Theorem \cite[Theorem 7.3]{book-NIT}, let $T=X+\alpha S$ for some $X$ that is continuously distributed according to $P_X$ and independent of $S$. Then, we have
\begin{subequations}
	\begin{align}
		\mathsf{C} &\ge \max_{P_X,\alpha} \mathsf{I}(T;Y)-\mathsf{I}(T;S) \\
		&=  \max_{P_X,\alpha} \mathsf{h}(T|S) - \mathsf{h}(T|Y) \\
		&= \max_{P_X,\alpha} \mathsf{h}(X) - \mathsf{h}(T|Y) \\
		&\overset{\text{(a)}}{\ge} \max_{P_X,\alpha,\hat{T}(Y)} \mathsf{h}(X) - \frac{1}{2}\log\Bigl(2\pi e \mathbb{E}\bigl[\bigl(T-\hat{T}(Y)\bigr)^2\bigr]\Bigr)\\
		&\overset{\text{(b)}}{\ge} \max_{P_X,\alpha} \mathsf{h}(X) - \frac{1}{2}\log\bigl(2\pi e {\rm LMMSE}(T,Y)\bigr) \\
		&\overset{\text{(c)}}{=} \max_{P_X} \mathsf{h}(X) - \frac{1}{2}\log\Bigl(\frac{2\pi e\mathbb{V}[X]\sigma^2}{\mathbb{V}[X] +\sigma^2} \Bigr) \label{eq:R0_temp}
	\end{align}
\end{subequations}
where (a) follows from \cite[Theorem 8.6.6]{book-EIT} and its corollary which provides an upper bound for conditional differential entropy $\mathsf{h}(T|Y)$ through the mean square error (MSE) of some optimal estimator $\hat{T}(Y)$; (b) follows by specifying the estimator $\hat{T}(Y)$ to be an linear minimum mean square error (LMMSE) estimator \cite{book-Kay1993}, which is suboptimal so that the MSE increases, and the resulting LMMSE is
\begin{subequations}
	\begin{align}
		&\quad{\rm LMMSE}(T,Y) \notag\\ 
		&=\mathbb{V}[T]-\frac{\mathbb{E}\bigl[ (T-\mathbb{E}[T])(Y-\mathbb{E}[Y]) \bigr]}{\mathbb{V}[Y]} \\
		&= \frac{(1-\alpha)^2\mathbb{V}[X]\mathbb{V}[S] + \sigma^2\mathbb{V}[X] + \alpha^2\sigma^2\mathbb{V}[S]}{\mathbb{V}[X]+\mathbb{V}[S]+\sigma^2}; \label{eq:lmmse}
	\end{align}
\end{subequations}
and (c) follows by letting $\alpha=\frac{\mathbb{V}[X]}{\mathbb{V}[X]+\sigma^2}$ which minimizes the ${\rm LMMSE}(T,Y)$ in \eqref{eq:lmmse} over $\alpha$.

Finally, we specify $X\sim{\rm Unif}([0,\mathsf{A}])$, so that $\mathsf{h}(X)=\log(\mathsf{A})$ and $\mathbb{V}[W]=\frac{\mathsf{A}^2}{12}$, then substitute $\mathsf{h}(X)$ and $\mathbb{V}[X]$ into \eqref{eq:R0_temp} leading to $R_0$ as shown in \eqref{eq:r0}. This ends the proof.
\end{proof}

\section{Proof of Theorem \ref{thm:r1}} \label{app:thm_R1}
\begin{proof}
From the equivalent channel formed by Scheme 1, we have the following achievable rate
\begin{subequations}
	\begin{align}
		\max_\alpha \mathsf{I}(T; Y') 
		& = \max_\alpha  \mathsf{h}(Y') - \mathsf{h}(\tilde{Z}'_\alpha\bmod \mathsf{A}_\Delta) \\
		& \overset{\text{(a)}}{=}   \log(\mathsf{A}_\Delta) - \min_\alpha \mathsf{h}( \tilde{Z}'_\alpha \bmod \mathsf{A}_\Delta) \\
		& \overset{\text{(b)}}{\ge}   \log(\mathsf{A}_\Delta) - \min_\alpha \mathsf{h}( \tilde{Z}'_\alpha ) \\
		& \overset{\text{(c)}}{\ge}  \log(\mathsf{A}_\Delta) - \min_\alpha h\bigl( \mathcal{N}(0, \mathbb{V}[\tilde{Z}'_\alpha])\bigr) \\
		& \overset{\text{(d)}}{=} \mathsf{R}_1,
	\end{align}
\end{subequations}
where (a) follows since $T+W_0\sim{\rm Unif}([0,\mathsf{A}_\Delta])$ implies that $Y'\sim{\rm Unif}([0,\mathsf{A}_\Delta])$ according to Crypto Lemma \cite[Lemma 1]{erez2004achieving}, 
(b) follows from the Modulo-Reduce-Entropy (MRE) Lemma \cite[Lemma A.3.2]{book-zamir2014lattice}, 
(c) follows since a Gaussian distribution maximizes the differential entropy under a variance constraint, and
(d) follows by solving the following minimization problem:
\begin{align}
	\alpha &= \arg\min_{\alpha}\mathbb{V}[\tilde{Z}'_\alpha] \notag \\
	&=\arg\min_{\alpha} \Bigl\{\frac{\Delta^2}{12}+\frac{(\alpha-1)^2\mathsf{A}(\mathsf{A}+2\Delta)}{12}+\alpha^2\sigma^2\Bigr\} \notag\\
	&=\frac{\mathsf{A}^2+2\mathsf{A}\Delta}{\mathsf{A}^2+2\mathsf{A}\Delta + 12\sigma^2},
\end{align}
where $\mathsf{R}_1$ is defined in \eqref{eq:r1}. This ends the proof.
\end{proof}

\bibliographystyle{IEEEtran}
\bibliography{ref_lib}

\begin{thebibliography}{10}
\providecommand{\url}[1]{#1}
\csname url@samestyle\endcsname
\providecommand{\newblock}{\relax}
\providecommand{\bibinfo}[2]{#2}
\providecommand{\BIBentrySTDinterwordspacing}{\spaceskip=0pt\relax}
\providecommand{\BIBentryALTinterwordstretchfactor}{4}
\providecommand{\BIBentryALTinterwordspacing}{\spaceskip=\fontdimen2\font plus
\BIBentryALTinterwordstretchfactor\fontdimen3\font minus
  \fontdimen4\font\relax}
\providecommand{\BIBforeignlanguage}[2]{{%
\expandafter\ifx\csname l@#1\endcsname\relax
\typeout{** WARNING: IEEEtran.bst: No hyphenation pattern has been}%
\typeout{** loaded for the language `#1'. Using the pattern for}%
\typeout{** the default language instead.}%
\else
\language=\csname l@#1\endcsname
\fi
#2}}
\providecommand{\BIBdecl}{\relax}
\BIBdecl

\bibitem{book-NIT}
A.~El~Gamal and Y.-H. Kim, \emph{Network Information Theory}.\hskip 1em plus
  0.5em minus 0.4em\relax Cambridge university press, 2011.

\bibitem{Costa1983}
M.~Costa, ``Writing on dirty paper (corresp.),'' \emph{IEEE transactions on
  information theory}, vol.~29, no.~3, pp. 439--441, 1983.

\bibitem{erez2005capacity}
U.~Erez, S.~Shamai, and R.~Zamir, ``Capacity and lattice strategies for
  canceling known interference,'' \emph{IEEE Transactions on Information
  Theory}, vol.~51, no.~11, pp. 3820--3833, 2005.

\bibitem{yu2001trellis}
W.~Yu and J.~M. Cioffi, ``Trellis precoding for the broadcast channel,'' in
  \emph{GLOBECOM'01. IEEE Global Telecommunications Conference (Cat. No.
  01CH37270)}, vol.~2.\hskip 1em plus 0.5em minus 0.4em\relax IEEE, 2001, pp.
  1344--1348.

\bibitem{yu2005trellis}
W.~Yu, D.~P. Varodayan, and J.~M. Cioffi, ``Trellis and convolutional precoding
  for transmitter-based interference presubtraction,'' \emph{IEEE Transactions
  on Communications}, vol.~53, no.~7, pp. 1220--1230, 2005.

\bibitem{weingarten2006capacity}
H.~Weingarten, Y.~Steinberg, and S.~S. Shamai, ``The capacity region of the
  gaussian multiple-input multiple-output broadcast channel,'' \emph{IEEE
  transactions on information theory}, vol.~52, no.~9, pp. 3936--3964, 2006.

\bibitem{mohseni2006capacity}
M.~Mohseni, \emph{Capacity of Gaussian Vector Broadcast Channels}.\hskip 1em
  plus 0.5em minus 0.4em\relax Stanford University, 2006.

\bibitem{Lapidoth-Moser-Wigger-2009}
A.~Lapidoth, S.~M. Moser, and M.~A. Wigger, ``On the capacity of free-space
  optical intensity channels,'' \emph{IEEE Transactions on Information Theory},
  vol.~55, no.~10, pp. 4449--4461, 2009.

\bibitem{chaaban2016capacity}
A.~Chaaban, Z.~Rezki, and M.-S. Alouini, ``On the capacity of the
  intensity-modulation direct-detection optical broadcast channel,'' \emph{IEEE
  Transactions on Wireless Communications}, vol.~15, no.~5, pp. 3114--3130,
  2016.

\bibitem{smith1971information}
J.~G. Smith, ``The information capacity of amplitude-and variance-constrained
  sclar gaussian channels,'' \emph{Information and control}, vol.~18, no.~3,
  pp. 203--219, 1971.

\bibitem{chan2005capacity}
T.~H. Chan, S.~Hranilovic, and F.~R. Kschischang, ``Capacity-achieving
  probability measure for conditionally gaussian channels with bounded
  inputs,'' \emph{IEEE Transactions on Information Theory}, vol.~51, no.~6, pp.
  2073--2088, 2005.

\bibitem{tomlinson1971new}
M.~Tomlinson, ``New automatic equaliser employing modulo arithmetic,''
  \emph{Electronics letters}, vol.~7, no.~5, pp. 138--139, 1971.

\bibitem{harashima1972matched}
H.~Harashima and H.~Miyakawa, ``Matched-transmission technique for channels
  with intersymbol interference,'' \emph{IEEE Transactions on Communications},
  vol.~20, no.~4, pp. 774--780, 1972.

\bibitem{erez2004achieving}
U.~Erez and R.~Zamir, ``Achieving 1/2 log (1+ snr) on the awgn channel with
  lattice encoding and decoding,'' \emph{IEEE Transactions on Information
  Theory}, vol.~50, no.~10, pp. 2293--2314, 2004.

\bibitem{mckellips2004simple}
A.~L. McKellips, ``Simple tight bounds on capacity for the peak-limited
  discrete-time channel,'' in \emph{International Symposium onInformation
  Theory, 2004. ISIT 2004. Proceedings.}\hskip 1em plus 0.5em minus 0.4em\relax
  IEEE, 2004, pp. 348--348.

\bibitem{book-EIT}
T.~M. Cover and J.~A. Thomas, \emph{Elements of Information Theory}.\hskip 1em
  plus 0.5em minus 0.4em\relax John Wiley \& Sons, 2012.

\bibitem{book-Kay1993}
S.~M. Kay, \emph{Fundamentals of statistical signal processing: estimation
  theory}.\hskip 1em plus 0.5em minus 0.4em\relax Prentice-Hall, Inc., 1993.

\bibitem{book-zamir2014lattice}
R.~Zamir, \emph{Lattice Coding for Signals and Networks: A Structured Coding
  Approach to Quantization, Modulation, and Multiuser Information
  Theory}.\hskip 1em plus 0.5em minus 0.4em\relax Cambridge University Press,
  2014.

\end{thebibliography}

\end{document}